\documentclass[reprint]{revtex4-1}
\usepackage{amsmath}
\usepackage{graphicx}
\usepackage{color}
\usepackage{amssymb}
\usepackage{wasysym}
\usepackage{amsthm}
\usepackage{soul}
\usepackage{mathtools}   

\input{epsf}

\begin{document}

\title{Entropy-driven formation of prolate and oblate cholesteric phases by computer simulations}

\author{Simone Dussi and Marjolein Dijkstra}
\affiliation{Soft Condensed Matter, Debye Institute for Nanomaterials Science, Utrecht University, Princetonplein 5, 3584 CC Utrecht, The Netherlands}

\begin{abstract}

Predicting the macroscopic chiral behaviour of cholesteric liquid crystals from the microscopic chirality of the particles is highly non-trivial, even when the chiral interactions are purely entropic in nature. Here we introduce a novel chiral hard-particle model, namely particles with a twisted polyhedral shape and obtain, for the first time, a stable fully-entropy-driven cholesteric phase by computer simulations. By slightly modifying the triangular base of the particle, we are able to switch from a left-handed prolate to a right-handed oblate cholesteric using the same right-handed twisted particle model. Furthermore, we find qualitative agreement with an Onsager-like theory, suggesting that the latter can be used as a quick tool to scan the huge parameter space associated to the microscopic chirality. Our results unveil how the competition between particle biaxiality and chirality is reflected into the nematic self-organization and new theoretical challenges on the self-assembly of chiral particles can be undertaken.

\end{abstract}

\maketitle
\section*{Introduction}
Since Onsager's prediction of a purely entropy-driven phase transition from an isotropic fluid  of infinitely long hard Brownian rods to an orientationally ordered nematic phase~\cite{onsager}, hard particles have  served as paramount models in condensed matter studies. The seminal work on crystallization of hard spheres revealed the crucial role of computer simulations in proving that order can be induced by entropy alone~\cite{alder,wood}. The macroscopic structure obtained by self-assembly of  colloidal particles is often directly   linked to the shape of the constituent building blocks~\cite{damasceno2012,vananders}. As soon as we move away from spherical particles a bewildering variety of thermodynamically stable structures with increasing complexity  arises ~\cite{damasceno2012,dijkstra}. As a result, a concurrent increase  of simulation studies  on hard particles  show that entropy  can be the sole driving force in the formation of crystals featuring different symmetries, plastic crystals, liquid crystals, and even quasi-crystals~\cite{damasceno2012,dijkstra,frenkelnem,frenkelsm,peroukidis,torquato,escobedo,kolli,damasceno2015,akbari}. Shape anisotropy is the essential ingredient to form liquid crystals (LC), phases featuring long-range orientational order but no or only partial positional order~\cite{degennes}. Hard bodies have been extensively employed also in the  field of LCs~\cite{mederos}. Thirty-five years after Onsager's prediction, the first entropy-stabilized nematic phase was observed in computer simulations of hard ellipsoids~\cite{frenkelnem}. In the nematic liquid-crystalline phase, the particles are on average aligned along a preferred  direction, identified by the nematic director $\mathbf{\hat{n}}$, but the positions are homogeneously distributed in the system. Additionally, hard spherocylinders were employed in simulations to demonstrate the thermodynamic stability of an entropy-driven smectic phase~\cite{frenkelsm}, in which the particles are orientationally ordered and arranged in smectic layers. This system has become a popular hard-particle model system to study LC phase behaviour~\cite{bolhuis,cuetos}. By introducing biaxiality in the  hard-particle shape, the long-searched biaxial nematic phase has also been  simulated~\cite{camp,peroukidis}. Furthermore, many other LC phases have been observed in simulations, which are entropy-driven, including  a cubic gyroid phase~\cite{ellison} and a twist-bend nematic phase~\cite{greco}. Surprisingly, from this long list of entropic LC and non-LC phases, a simulation evidence of a cholesteric phase made of hard particles is still missing, despite the facts that it was the first LC phase experimentally discovered~\cite{reinitzer} and that an entropic cholesteric phase was already theoretically predicted forty years ago~\cite{straley}. A cholesteric phase  displays an helical chiral arrangement of the director field, $\mathbf{\hat{n}}(z)= \left\{ \cos \left(\frac{2\pi}{\mathcal{P}} z\right), \sin \left( \frac{2\pi}{\mathcal{P}} z\right), 0\right\}$, with $z$ the axis of the macroscopic twist (chiral director) and $\mathcal{P}$ the cholesteric pitch that determines the typical length scale associated to the helical periodicity. Several theoretical studies have been dedicated to better understand the link between microscopic and macroscopic chirality. A unified picture has still yet to be achieved since it is clear that the cholesteric pitch $\mathcal{P}$ depends in a non-trivial way on both the single-particle properties and the thermodynamic state of the system (for example see~\cite{straley,harris1997,harris1999,tombolato2006,wensink2009,belli}). The microscopic origin of chirality has also been the focus of experimental studies on colloidal systems~\cite{tombolato2006,grelet,zanchetta}, and of computer simulation studies based on  strongly  chiral attractive interactions~\cite{memmer,varga2006,melle}. The main reason that a cholesteric phase of hard bodies has never been observed in simulations is due to the fact that the cholesteric pitch length is on the order of hundreds or thousands times the particle length, and that huge system sizes, beyond our computational limits, are needed to accomodate the cholesteric pitch. Recently, hard helices have been introduced as a simple particle model, but the formation of a cholesteric phase has never been addressed in simulations as the focus of these studies was more on the intriguing chiral phases that occur at high densities~\cite{kolli}. Hence it is still unsettled if and how a twist in the particle shape gives rise to the cholesteric order and several questions, that have been addressed by computer simulations for the achiral nematic phase, like nucleation, wetting etc...~\cite{cuetos,dijkstra2001}, remained so far unexplored for the cholesteric phase.

\section*{Results}
\subsection*{A novel chiral hard particle model}
\begin{figure}[h!t]
\center
\includegraphics[width=0.95\linewidth]{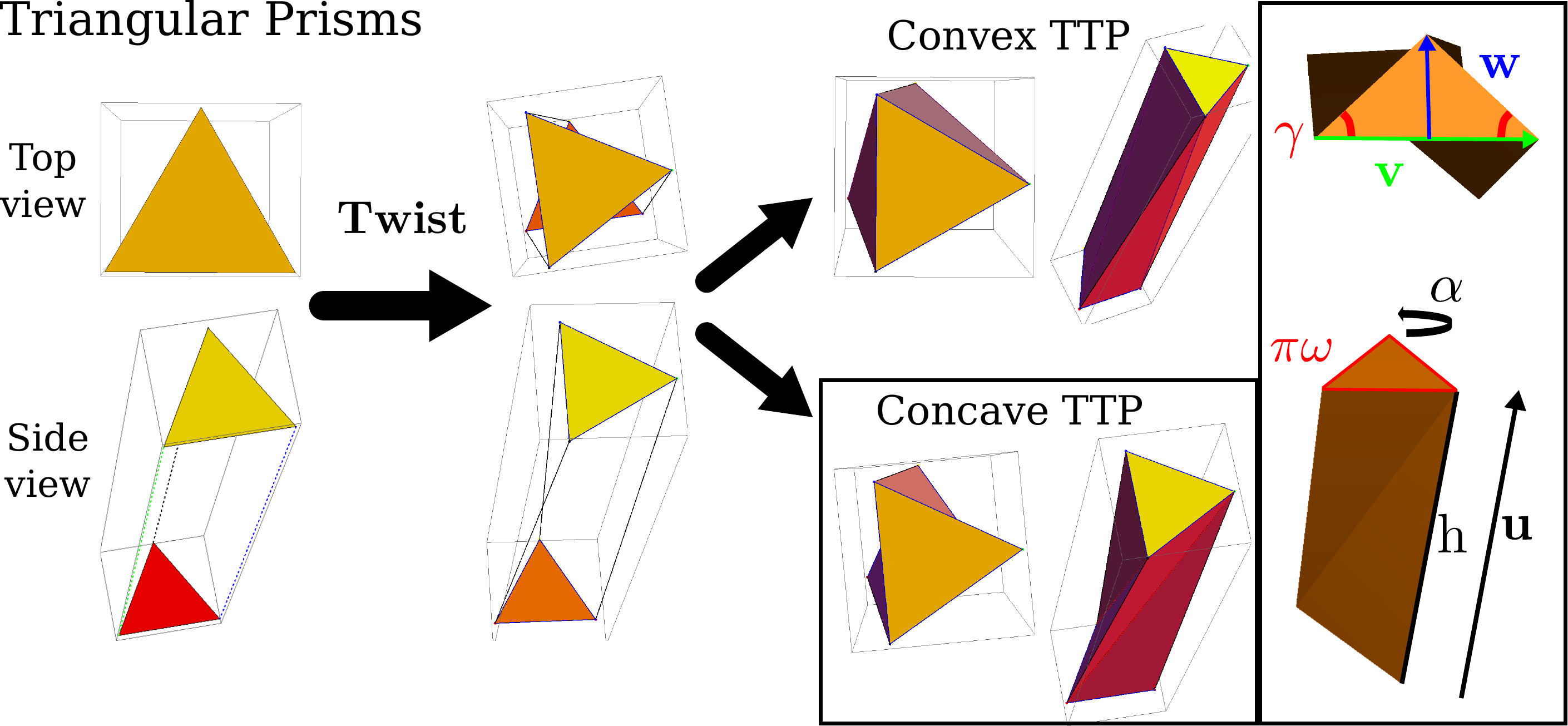}
\caption{\label{fig1}A twisted triangular prism (TTP) is constructed from an elongated prism of height $h$ with (isosceles) triangular bases, determined by the angle $\gamma$, and perimeter $\pi \omega$. The width $\omega$ is used as the unit of length. To introduce chirality, one triangular base is twisted by an angle $\alpha$ relative to the other one and additional edges are constructed to obtain flat faces. The orientations of the main axes are described by the unit vectors $\mathbf{\hat{u}}$ (long), $\mathbf{\hat{v}}$ (medium) and $\mathbf{\hat{w}}$ (short).} 
\end{figure}

Here, we show the first fully-entropic cholesteric phase obtained by computer simulations of a novel type of particle, namely \emph{hard twisted polyhedra}. As we explain below, this particle model presents several shape features that can be easily tuned, e.g. aspect ratio, convexity, biaxiality, handedness, degree of twist (or molecular pitch), number of polyhedral faces. A systematic study of how these properties, some of which are intuitively associated to microscopic chirality and liquid-crystalline behaviour, affect the self-assembly of many of such particles,  can be efficiently carried out by perfoming computer simulations. In particular, here  we study the nematic phase behaviour of the simplest shape of this class, i.e., twisted triangular prisms (TTP). Our particle is obtained by twisting one base of an elongated triangular prism of  height $h$ by an angle $\alpha$   relative to the other base and by adding additional edges to ensure flat faces (see Fig.~\ref{fig1}). Remarkably, depending on the choice of which vertices are connected by these additional edges, it is possible to build both concave and convex chiral particles. The triangular base has fixed perimeter $\pi \omega$, such that in the limit of infinite number of sides (circle) the width $\omega$ coincides with particle diameter. In this study we consider concave TTPs with either equilateral or isosceles triangular bases defined by the base angle $\gamma$. When the top triangular base is rotated clockwise the twist angle $\alpha$ is positive and it is tempting to call the TTP right-handed. We return to this definition of particle handedness when we discuss our results for the oblate cholesteric phases. We note that $\alpha$ should be less than or equal to the smallest angle of the base to avoid self-intersection of the particle shape. To detect overlaps between particles, i.e., the key ingredient in Monte Carlo (MC) simulations aimed to study the self-assembly of hard particles, we use an algorithm based on triangle-triangle intersection detection~\cite{rapid}, which is also  suitable  for concave shapes. Analogously to spherocylinders (and other hard-rod models), the nematic phase  can be stabilized at sufficiently high aspect ratio ($h/\omega$)~\cite{bolhuis}, whereas the particle chirality can be tuned by changing the twist angle $\alpha$ that also changes the molecular pitch $p \simeq 2\pi h/\alpha$. Additionally, by further modifying the particle shape (changing the base) we study how the competition between biaxiality and chirality propagates from microscopic (single-particle) to macroscopic (self-assembled structure) level.

Nematic phases formed by biaxial particles can be distinguished in prolate ($N_+$), oblate ($N_-$) and biaxial ($N_b$), depending on which particle axes feature long-range orientational order~\cite{straleybiaxial}.
By defining a shape parameter based on the lengths of the particle main axes $\nu=h/m-m/s$ ($h$=height, $m$=medium, $s$=short), a prediction of which nematic occurs can be made.
For $\nu>0$,  the long axes $\mathbf{\hat{u}}$ of the particles are aligned in the nematic phase along a common director, resulting into a $N_+$ phase, whereas for $\nu<0$ the short particle axes  $\mathbf{\hat{w}}$ are aligned and a $N_-$ phase is formed. When $\nu \sim 0$ a biaxial phase can be stabilized, provided that other conditions that are strongly shape-dependent are also satisfied (for example for rounded board-like particles considered in Ref.~\cite{peroukidis} there is an additional condition of $h/s \geq 9$).

\subsection*{Formation of the cholesteric phase}
To investigate whether or not the twist in the particle shape is transmitted at a macroscopic level, we perform MC simulations of thousands of TTPs using different initial configurations and boundary conditions, finding consistent results. In this section we present results from MC simulations using standard periodic boundary conditions (PBC) in the $NPT$ ensemble, i.e., at fixed number of particles $N$, pressure $P$, and temperature $T$,  and we investigate the kinetic pathways for the formation of a {\em prolate} cholesteric phase.

\begin{figure*}[htb]
\center
\includegraphics[width=0.95\linewidth]{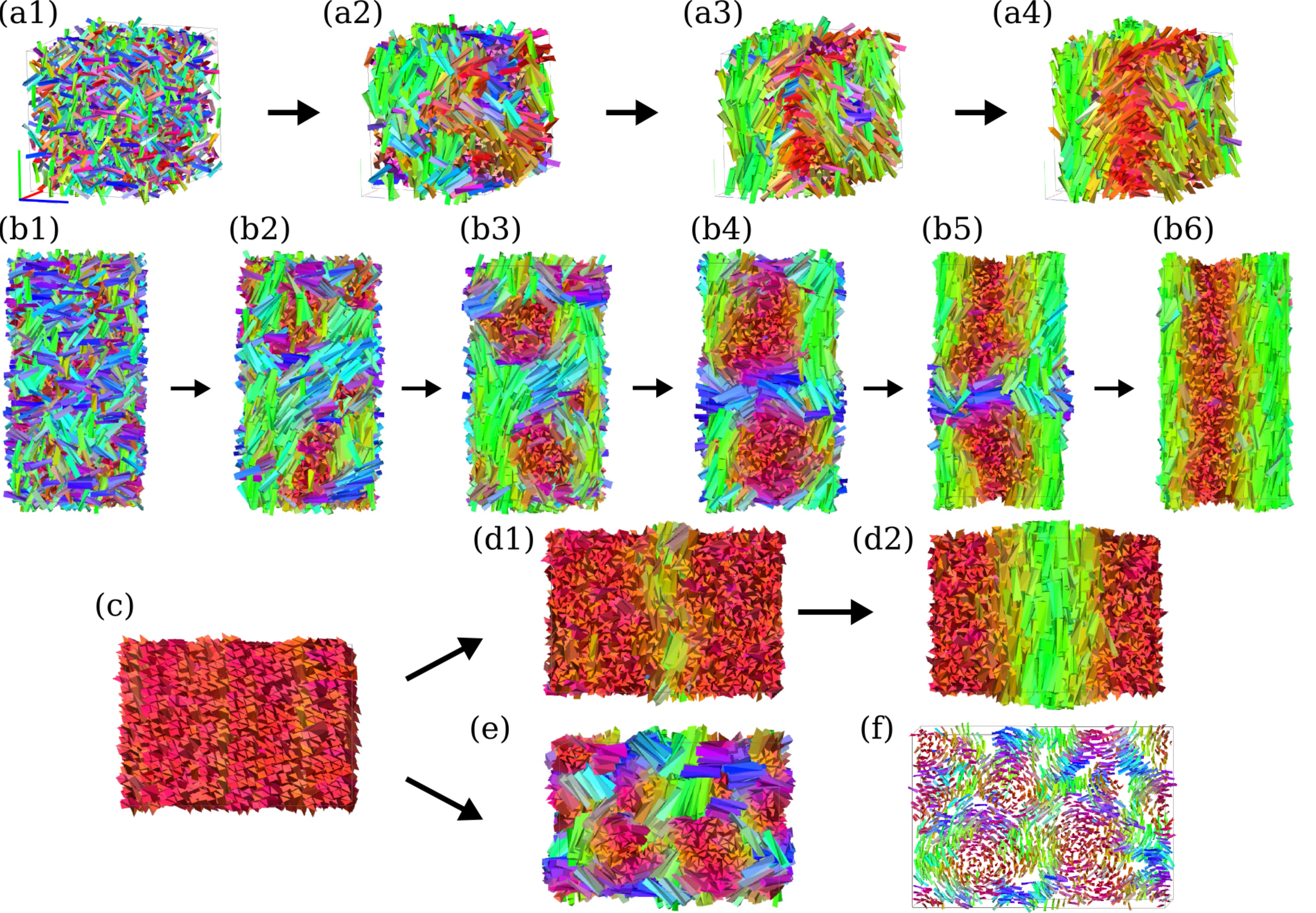}
\caption{\label{fig2} Kinetic pathways for the formation of a prolate cholesteric phase from MC simulations in the $NPT$ ensemble under standard periodic boundary conditions. The particles are colored according to the orientation of their main axis $\mathbf{\hat{u}}$. {\bf (a)} Snapshot time series of $2000$ TTP with aspect ratio $h/\omega=5$, base angle $\gamma=1.0$, and particle twist angle $\alpha=0.7$, starting from an isotropic state (a1) and imposing a pressure $\beta P\omega^3=1.5$ (see also Supplementary Movie 1). Evolution in time corresponds to an increase of the system packing fraction. The resulting phase (a4) is a \emph{left-handed} cholesteric. {\bf (b)} Snapshot time series of $3200$ TTP with $h/\omega=5$, $\gamma=0.75$, and $\alpha=0.7$, starting from an isotropic state (b1) and imposing $\beta P\omega^3=1.9$ (see also Supplementary Movie 3). In this case a spinodal instability is observed: chiral nematic domains are first formed, resembling a blue phase (b3), and slowly merge into a cholesteric defect-free phase. {\bf (c)} Dense aligned state of $2304$ TTP with $h/\omega=5$, $\gamma=0.75$, and $\alpha=0.7$ used as initial configuration for expansion runs. {\bf (d)} At pressure $\beta P \omega^3=2.1$ the twist slowly propagates from the center of the system (d1) to the entire system (d2) and also in this case a cholesteric phase is formed (see also Supplementary Movie 4). {\bf (e)} At higher pressure ($\beta P\omega^3=2.9$) an instability with respect to nematic orientational fluctuations as well as smectic layering fluctuations is observed  (see also Supplementary Movie 5), resulting into a \emph{metastable} state of chiral domains composed of highly aligned particles (e), which bears close resemblance to a blue phase. See also panel {\bf (f)} where the particle size is reduced. }
\end{figure*}

\begin{figure*}[htb]
\center
\includegraphics[width=0.95\linewidth]{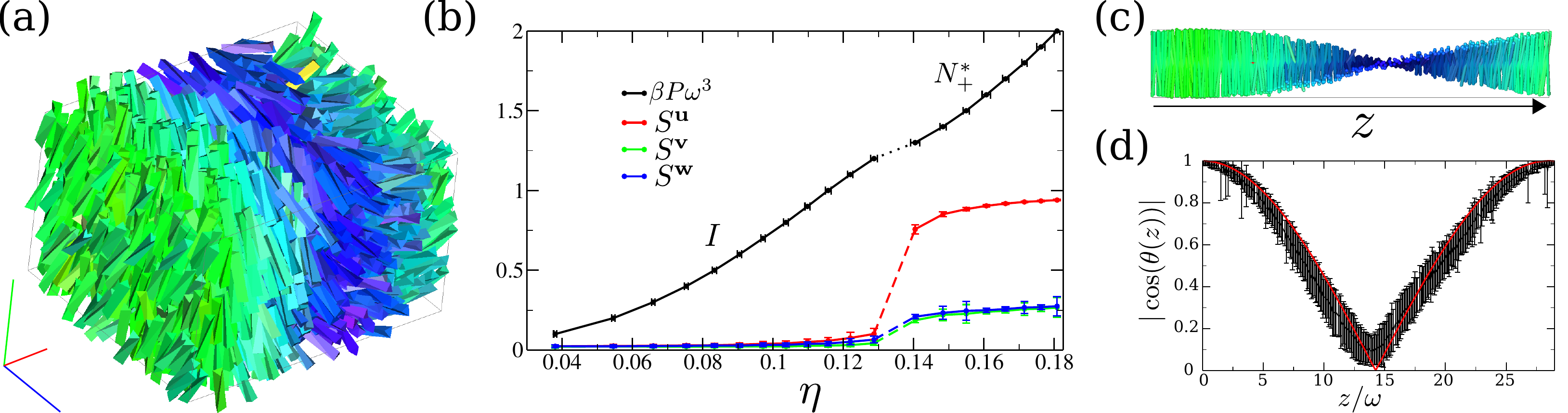}
\caption{\label{fig3} {\bf (a)} Typical configuration as obtained from simulations of a prolate cholesteric phase of TTPs with aspect ratio $h/\omega=5$, twist angle $\alpha=0.7$, and base angle $\gamma=0.75$ using PBC. Particles are colored according to the orientation of their main axis $\mathbf{\hat{u}}$. {\bf (b)} Equation of state and nematic order parameters $S$ associated to the particle main axis $\mathbf{u}$,$\mathbf{v}$,$\mathbf{w}$, as a function of the packing fraction $\eta$. {\bf (c)} Nematic director profile $\mathbf{\hat{n}}_{\mathbf{\hat{u}}}$ depicted by using rods color-coded according to their orientation. {\bf (d)} The twist is quantified by averaging hundreds of such profiles after measuring $|\cos(\theta(z))|$, with $\theta$ the twist angle along the $z$-direction that identifies the chiral director. The fit used to extract the cholesteric pitch $\mathcal{P}$ is indicated with a red line (see text and Methods for more details).} 
\end{figure*}

{\em Left-handed cholesteric phase.} We first consider $2000$ TTPs  with a strong particle twist $\alpha=0.7$, and with an aspect ratio $h/\omega=5$ with an almost equilateral base $\gamma=1.0$, yielding $\nu \simeq 3.25>0$. We start from an isotropic phase (I) and perform a compression by fixing the pressure $\beta P \omega^3 =1.5$ with $\beta=1/(k_B T)$ and  $k_B$ Boltzmann's constant. Fig.~\ref{fig2}(a) clearly demonstrates the formation of a prolate nematic phase ($N_+^*$) with a spontaneous macroscopic \emph{left-handed} twist upon increasing the density. The resulting structure has been characterized using appropriate order parameters as shown in Fig.~\ref{fig3} and described in detail in the following section. We notice here that the \emph{opposite} handedness of the cholesteric phase with respect to the particle twist is consistent with theoretical predictions~\cite{straley,harris1997,belli,dussi}  for chiral particles with a large molecular pitch $p$, in this case $p/\omega \simeq 44.8$. The phase transformation from I to $N_+^*$ is driven first by  nematic fluctuations due to the anisotropy of the overall particle shape, and subsequently, the nematic phase becomes twisted as a result of the finer details of the chiral particle shape. Additionally, we confirm the stability of the $N_+^*$ phase by starting from a uniaxial nematic state.  We observe that the achiral order is clearly unstable since a twist starts to propagate slowly throughout the whole system as shown in the Supplementary Movie 2.   
Our simulations show that a stable prolate cholesteric phase is found for a large range of twist angles, i.e., $0 \lesssim \alpha \lesssim \gamma$, and base angles $0.55 \lesssim \gamma \lesssim \pi/3$. 

{\em Kinetic pathways.} We now investigate in more detail the kinetic pathways leading to $N_+^*$ in case of short chiral rods. To this end, we focus on large systems of TTPs with aspect ratio $h/\omega=5$, an isosceles base with angle $\gamma=0.75$ and a particle twist angle $\alpha=0.7$, yielding a shape parameter $\nu \simeq 1.62>0$. First, we determine the pressure $\beta P \omega^3$ and nematic order parameters $S$ associated to the three main axes $\mathbf{u},\mathbf{v},\mathbf{w}$ as a function of packing fraction $\eta$ as shown in Fig.~\ref{fig3}(b). From the equation of state, we find that the $IN_+^*$ transition occurs at a pressure $\beta P \omega^3 \simeq 1.25$, and the transition from a cholesteric phase to a higher ordered one, that we generically denote as chiral smectic ($Sm^*$), takes place at a pressure  $\beta P \omega^3 \simeq 2.6$ (see Supplementary Fig.~1 for more details). A closer look to the formation of the cholesteric phase reveals that for sufficiently high supersaturation of the isotropic phase ($\beta P \omega^3 =1.9$), the transformation proceeds via spinodal decomposition, in analogy with achiral short spherocylinders~\cite{cuetos}, see Fig.~\ref{fig2}(b). We clearly observe that the system is unstable as immediately many small nematic clusters with different orientations are formed  throughout the system, which subsequently start to twist. Interestingly, the intermediate phase looks remarkably similar to a blue phase~\cite{degennes,melle}, but after a long equilibration time the twisted nematic domains start to merge  and the system relaxes to a cholesteric phase. Finally, we also study the phase transformation starting from a dense aligned phase. We perform $NPT$-MC simulations both at a pressure $\beta P \omega^3 =2.1$ corresponding to a stable $N_+^*$ phase, and at  $\beta P \omega^3 =2.9$, where the $Sm^*$ phase is expected to be stable. For $\beta P \omega^3 =2.1$, we indeed find that the director field immediately starts to twist in the aligned phase of TTPs, resulting into a cholesteric phase as shown in Fig.~\ref{fig2}(c)-(d). On the other hand, at $\beta P \omega^3 =2.9$, the system is unstable with respect to both nematic orientational fluctuations as well as smectic layering fluctuations as seen in Fig.~\ref{fig2}(e), and again the final structure bears close resemblance to a blue phase, which corresponds to a metastable (kinetically arrested) state as the $Sm^*$ phase is expected to be the stable state. However, this suggests an intriguing competition between packing and chirality at high pressures that will be further investigated in future studies.

\begin{figure*}[htb]
\center
\includegraphics[width=0.95\linewidth]{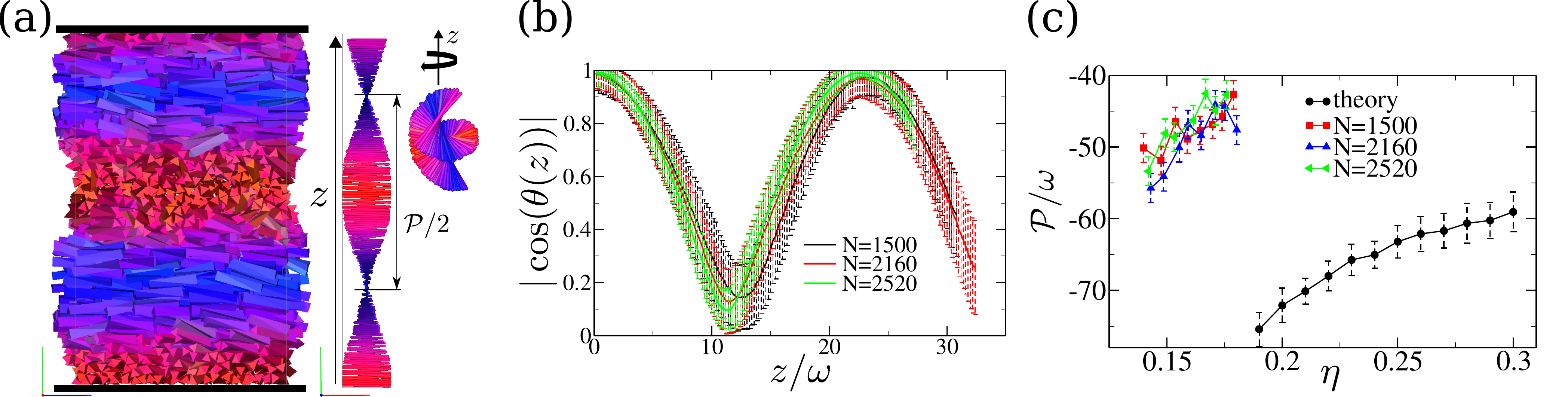}
\caption{\label{fig4} {\bf (a)} Typical left-handed cholesteric along with cartoons of $\mathbf{\hat{n}}_{\mathbf{\hat{u}}}$ as obtained from $NVT$-MC simulations on systems of TTPs with aspect ratio $h/\omega=5$, twist angle $\alpha=0.7$, and base angle $\gamma=0.75$ confined between two planar hard walls (in the $z$-direction). The particles are colored according to the orientation of their main axis $\mathbf{\hat{u}}$. {\bf (b)} Local director profiles $\theta(z)$ obtained by simulations using varying number of particles $N$ and varying box sizes but at fixed packing fraction $\eta \simeq 0.16$. {\bf (c)} Cholesteric pitch $\mathcal{P}$ versus packing fraction $\eta$ as obtained from simulations (using hard walls) and a second-virial density functional theory~\cite{belli,dussi}.} 
\end{figure*}

\subsection*{Equilibrium cholesteric pitch}
We now turn our attention to the helical structure of the cholesteric phase. We want to determine the equilibrium cholesteric pitch $\mathcal{P}$ and study how boundary conditions and finite-size effects influence this quantity, that ultimately is the parameter quantifying the macroscopic chirality. To this end, we analyze the spatial dependence of the nematic director by dividing the system in slabs~\cite{varga2006,melle} along the axis of the macroscopic twist, also called chiral director, which is oriented along the $z$-axis. We then compute the nematic director $\mathbf{\hat{n}}_{\mathbf{\hat{u}}}(z)$ corresponding to the long particle axis $\mathbf{\hat{u}}$, an example of such a profile is depicted in Fig.~\ref{fig3}(c), and the associated nematic order parameter $S^{\mathbf{u}}$ as displayed in Fig.~\ref{fig3}(b). The same procedure is repeated for the medium particle axis $\mathbf{\hat{v}}$ and the short axis $\mathbf{\hat{w}}$ to obtain $S^{\mathbf{v}}$, $S^{\mathbf{w}}$ (see Fig.~\ref{fig3}(b)), and $\mathbf{\hat{n}}_{\mathbf{\hat{v}}}$, $\mathbf{\hat{n}}_{\mathbf{\hat{w}}}$ (not shown). The twist is quantified by averaging $|\cos(\theta(z))| \equiv |\mathbf{\hat{n}}_{\mathbf{\hat{u}}}(z) \cdot \mathbf{\hat{n}}_{\mathbf{\hat{u}}}(z=0)|$ over hundreds of independent configurations and performing a one-parameter fit using $|\cos \left(2\pi z/\mathcal{P} \right)|$ to extract the cholesteric pitch $\mathcal{P}$, shown with red lines in Fig.~\ref{fig3}(d). In addition, we compute orientational pair-correlation functions along the chiral director, as introduced in Ref.~\cite{memmer}, to confirm the helical structure and the sense of the macroscopic twist (see Methods and Supplementary Fig.~2 for more details).

It is important to note that in the case of PBC the nematic director should be the same at the edges of the simulation box, i.e., $\cos(\theta(z=L_z))=1$ in Fig.~\ref{fig3}(d) with $L_z$ the box length in the $z$-direction. As a consequence, the cholesteric pitch $\mathcal{P}$ must be commensurate with $L_z$, i.e., $L_z$ should be at least  $\sim \mathcal{P}/2$ to observe a twist in the nematic phase~\cite{memmer}. By allowing the box shape to relax, either by performing $NPT$ simulations or $NVT$ simulations using a variable box shape, we expect the accuracy of the equilibrium pitch measurement to improve, but by repeating our simulations for different system sizes,  we still observed a dependence on the initial box size (see Supplementary Fig.~3(a)).

In order to circumvent the commensurability problems with pitch and box size, we embed the system  between two planar hard walls in such a way that the nematic director can freely choose its orientation at both walls, and we perform simulations in the $NVT$ ensemble.  As can be observed from Fig.~\ref{fig4}(a)-(b), the nematic director profile is indeed  not commensurate anymore with $L_z$ thereby allowing for a full relaxation of the macroscopic chiral twist. Since we simulate sufficiently large system sizes at state points that are well inside the stable region of the cholesteric phase, we expect that surface effects, such as pronounced layering or increased biaxiality~\cite{vanroij,dijkstra2001}, should be negligible. We indeed observe that the walls only affect the structure at  distances smaller than $\sim$ one particle diameter from the wall (see Supplementary Fig.~4). In order to support this, we determine the equilibrium pitch  $\mathcal{P}$  using different system sizes (different number of particles and different box dimensions), but at fixed packing fraction $\eta$. Panels (b) and (c) of Fig.~\ref{fig4} show indeed consistent results for the nematic director profile as well as for the value of $\mathcal{P}$. We thus regard this method to be the most convenient and reliable way for calculating the equilibrium pitch $\mathcal{P}$, in analogy with the conclusions of Ref.~\cite{varga2006}.   

Finally, we perform simulations using twisted boundary conditions (TBC)~\cite{allen}. We find good equilibration of our cholesteric phases of TTPs, as evidenced in Supplementary Fig.~3(b) by the much smaller errror bars on the nematic director profile and the difference of $\pi/2$ in the $\theta$ angle at the edges of the box as imposed by the TBC. However, the use of TBC may result into an over- or undertwisted cholesteric phase, and only by measuring the pressure tensor, which is unfortunately not straightforward for hard particles, it would be possible to extract the equilibrium value of $\mathcal{P}$~\cite{frenkeltwist}. This procedure is based on a quadratic approximation around the free-energy minimum~\cite{frenkeltwist} and TTPs will be a suitable system to test this approach.

\subsection*{Comparison with second-virial theory}
\begin{figure}[h!t]
\center
\includegraphics[width=0.95\linewidth]{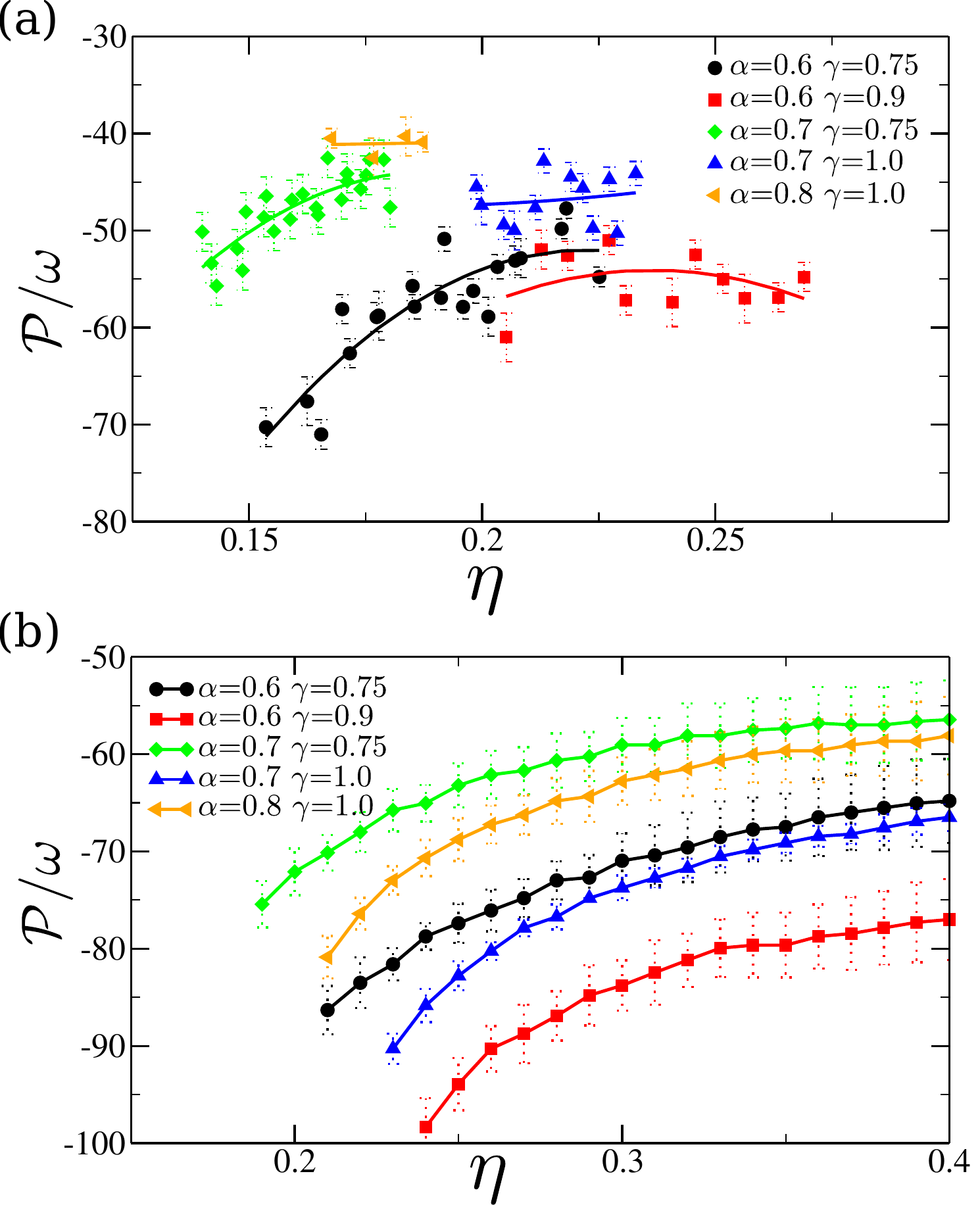}
\caption{\label{fig5} {\bf (a)} Cholesteric pitch $\mathcal{P}$ versus packing fraction $\eta$ as obtained by MC simulations (using hard walls) for TTPs with varying twist angles $\alpha$ and base angle $\gamma$ as labeled. Lines are polynomial fits used as guides to the eye. {\bf (b)} Theoretical predictions for the same particle models obtained by second-virial DFT~\cite{belli,dussi}. An increase in the twist angle $\alpha$ while keeping fixed the base angle $\gamma$ corresponds to a shorter cholesteric pitch $|\mathcal{P}|$. Analogously, keeping fixed $\alpha$ and decreasing $\gamma$ enhance the particle chirality and a corresponding shorter $|\mathcal{P}|$. However, this effect seems to be overestimated by  theory resulting in a slightly different chiral ranking for the models considered  here. In Supplementary Fig.~5(a) the same results are shown by using the nematic order parameter instead of the packing fraction and similar conclusions are drawn. } 
\end{figure}

The availability of cholesteric phases obtained from particle-based simulations provides a new testing ground for the theoretical framework describing the chiral organization in liquid crystals. We apply the recently developed second-virial density functional theory (DFT)~\cite{belli,dussi} to our system of TTPs and calculate the density dependence of $\mathcal{P}$. Our DFT is an extension of Onsager's theory~\cite{onsager} corrected with a  Parsons-Lee (PL) prefactor to deal with the finite size of the particles~\cite{parsons,lee}. It represents an advancement over Straley's approach~\cite{straley} as it does not consider the chirality in a perturbative way and it is combined with a MC integration to make it suitable for a wide range of particle models~\cite{dussi}. A detailed description can be found in Refs.~\cite{belli,dussi}. In Fig.~\ref{fig4}(c) we present our results as obtained from simulations along with the DFT predictions. We plot  $\mathcal{P}$ as a function of $\eta$ in the range where the cholesteric phase is stable. We observe that the theory correctly captures the twist handedness, the magnitude and the trend of $\mathcal{P}$ as a function of $\eta$. In addition, we study the effect of  particle shape on the cholesteric pitch  $\mathcal{P}$. In  Fig.~\ref{fig5}(a), we present simulation results for the pitch $\mathcal{P}$ as a function of $\eta$ for varying twist angle $\alpha$ and base angle $\gamma$. Comparing the results for $\alpha=0.6$ (red curve) with $\alpha=0.7$ (green) at fixed $\gamma=0.75$, or  $\alpha=0.7$ (blue) with $\alpha=0.8$ (yellow) at  $\gamma=1.0$, reveals that $|\mathcal{P}|$ decreases upon increasing $\alpha$, i.e., increasing the microscopic chirality of the particle, as expected. Analogously,  by decreasing the base angle $\gamma$, the surface associated with the longer side of the base gets larger, which effectively increases the particle chirality, thereby yielding a smaller pitch $|\mathcal{P}|$. This trend can be appreciated by comparing the results for  $\gamma=1.0$ (blue) with $\gamma=0.75$ (green) at fixed $\alpha=0.7$ or  $\gamma=0.9$ (red) with $\gamma=0.75$ (black) at fixed $\alpha=0.6$. Despite an overall little underestimation of the macroscopic twist, Fig.~\ref{fig5}(b) shows that all these trends are well-captured by our DFT calculations: increasing the particle chirality, by either twisting the particle more (increasing $\alpha$) or increasing the particle biaxiality (by decreasing $\gamma$), results indeed in a smaller cholesteric pitch  $|\mathcal{P}|$. However, we notice that the effect of decreasing $\gamma$ on particle chirality is overestimated by the DFT with  respect to the results obtained from simulation of the many-particle system. Nevertheless, the DFT can be used as a reliable and quick tool for predicting the macroscopic chiral behaviour from the microscopic chiral particle properties. We therefore use our theory to study the effect of TTPs with multiple twists on the cholesteric pitch $P$.  Our DFT calculations as shown in Supplementary Fig.~5(b) reveal that upon decreasing the microscopic pitch length $p/\omega$, the sense of the macroscopic twist changes from opposite to same handedness with in between a regime where a twist inversion occurs  with packing fraction, which is analogous to previous results on hard helices~\cite{belli,dussi}. In the conclusions we will discuss possible improvements for the theoretical framework.

\subsection*{Right-handed oblate cholesteric phase}
\begin{figure}[h!tb]
\center
\includegraphics[width=0.95\linewidth]{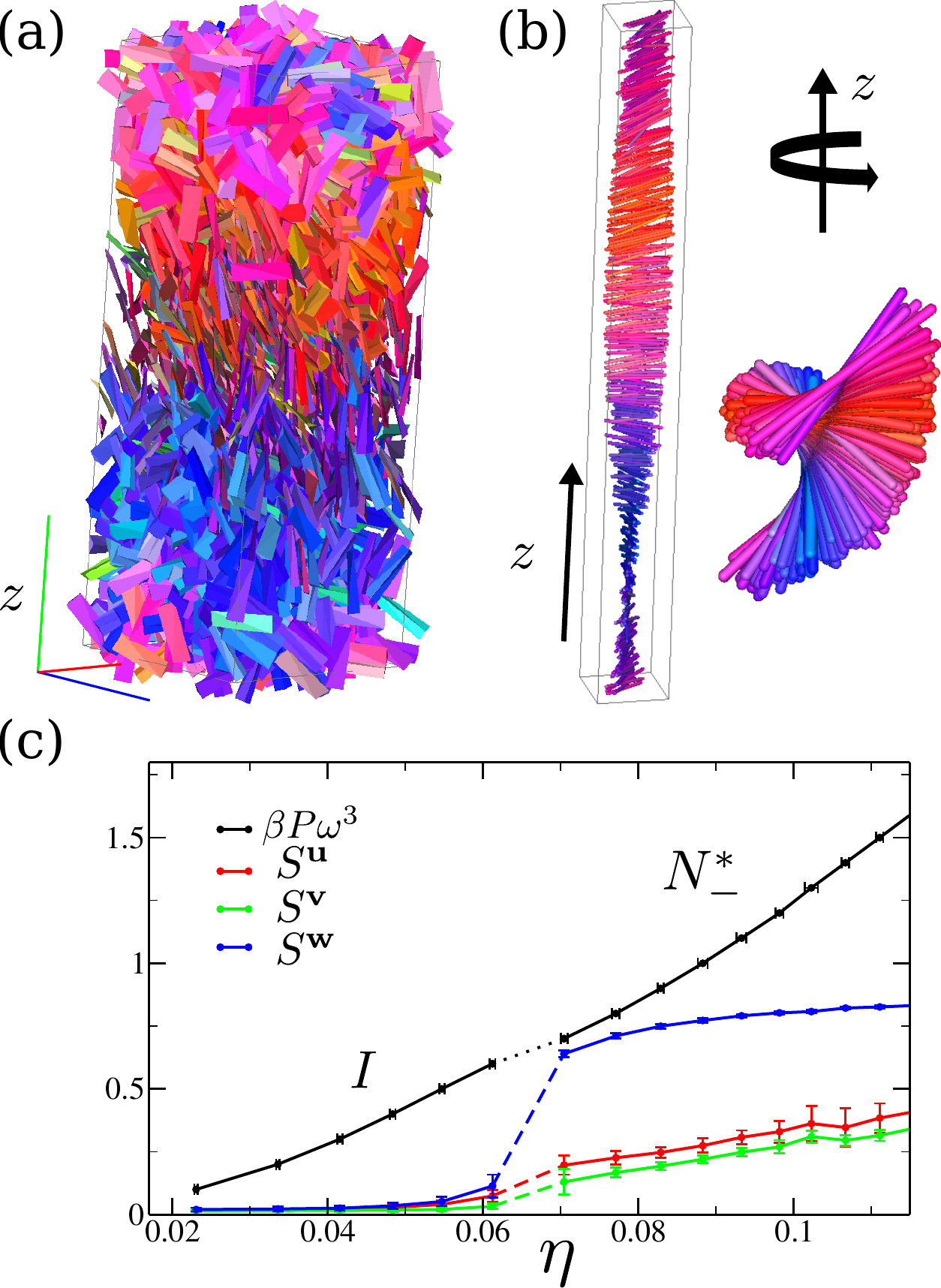}
\caption{\label{fig6} {\bf (a)} Oblate cholesteric phase of TTPs with $h/\omega=5$, $\alpha=0.4$, $\gamma=0.4$ as obtained from $NPT$-MC simulations ($\beta P \omega^3=1.0$) using PBC. Particles are colored according to the orientation of their \emph{short} axes $\mathbf{\hat{w}}$. {\bf  (b)} Cartoon of the nematic director $\mathbf{\hat{n}}_{\mathbf{\hat{w}}}$ exhibiting a \emph{right-handed} twist. {\bf (c)} Equation of state and nematic order parameters  $S^{\mathbf{u}}$, $S^{\mathbf{v}}$, $S^{\mathbf{w}}$ as a function of  packing fraction $\eta$, confirming that the transition is specular to the isotropic-prolate cholesteric phase transition (cfr. Fig.~\ref{fig3}(b)).} 
\end{figure}

Finally, we further modify the particle shape by decreasing the base angle $\gamma$ while keeping the aspect ratio $h/\omega=5$ fixed. In this way, we construct TTPs with shape parameter $\nu<0$, which should stabilize an oblate nematic phase. Indeed, our simulations reveal  the formation of an oblate nematic phase ($N_-^*$) with a helical chiral arrangement of the local nematic director field corresponding to the short particle axis $\mathbf{\hat{w}}$ as exemplarily shown in Fig.~\ref{fig6} for TTPs  with $h/\omega=5$, $\gamma=0.4$, $\alpha=0.4$, yielding $\nu \simeq -1.41$.  We observe that the orientation of the long particle axis $\mathbf{\hat{u}}$ is isotropic whereas the nematic director $\mathbf{\hat{n}}_{\mathbf{\hat{w}}}$ associated to the short axis $\mathbf{\hat{w}}$ displays the expected helical structure. Surprisingly, the macroscopic twist is now \emph{right-handed} in contrast with the left-handed twist as observed for the prolate cholesteric phase of the same particle model but with a different $\alpha$ and $\gamma$ (Fig.~\ref{fig3}), which seems to be counterintuitive. However, this can be explained as follows. Despite the fact that the twist angle $\alpha>0$, meaning that the particle is twisted in a right-handed fashion along the long particle axis $\mathbf{\hat{u}}$, it also corresponds to a left-handed twist in the short particle axis $\mathbf{\hat{w}}$. As only the short particle axes show  orientational order in an oblate nematic phase, and the particles are weakly chiral, we expect a macroscopic twist  that is opposite to that of the short axis, i.e., a right-handed macroscopic twist, as indeed observed in our simulations.  The Supplementary Movie 6 (see also equation of state and order parameters in Fig.~\ref{fig6}(c) ) shows that the I-$N_-^*$ phase transformation  is specular to that of the prolate cholesteric phase. 

\section*{Discussion}
In conclusion, we have shown by computer simulations of twisted triangular prisms that entropy alone can stabilize both oblate and prolate chiral nematic phases. Our results showcase once more that attributing uniquely a value to the microscopic chirality is not trivial. In this simple model we need to combine the value of the twist angle $\alpha$, shape parameter $\nu$ and the microscopic pitch $p$ to predict the sense of the twist. A more complicated competition between biaxiality and chirality is expected when $\nu \sim 0$ and a biaxial nematic phase should occur. Since a double twist cannot be space-spanning, it is interesting to investigate if the macroscopic chirality disappears or another stabilization mechanism comes into play.
Twisted polyhedra will be useful models to address this and other fundamental questions. For example, preliminary simulations on mixtures of particles with different handedness show that racemic mixtures form achiral nematic phases, as expected from theory~\cite{degennes}, indicating that chirality alone is not enough to drive phase separation in systems of hard particles and size asymmetry is required. By considering also depletant particles it will be possible to gain novel insights in experiments where entropy, chirality and depletion are the dominant forces~\cite{sharma,kang}.

In addition, our simulations show qualitative agreement with theoretical predictions from an Onsager-like DFT, thereby providing confidence that the theory yields reliable results and can thus serve as  a guide for future studies.  For example, to study nucleation and growth of  cholesteric phases, addressing questions like how the chirality changes the shape of the nematic nucleus, longer particles are needed~\cite{cuetos} for which $\mathcal{P}$ is expected to be larger and therefore a careful choice of the shape is essential. However, it is also evident that the Parsons-Lee (PL) correction does not rescale the packing fraction of twisted polyhedra accurately enough and overestimates the IN* transition. This may be remedied by a better rescaling factor than the PL correction, or by a more accurate microscopic theory such as fundamental measure theory~\cite{marechal}. 
The DFT also overestimates $\mathcal{P}$ compared to simulations, i.e., it underestimates the macroscopic chiral twist of a N* phase. A similar conclusion was also drawn in previous work on  attractive chiral spherocylinders~\cite{varga2006}. These issues need further investigations. 

Finally, recent advancements in chemical synthesis of nanocrystals with polyhedral shape~\cite{xia} and the use of polarized light to introduce chirality in the shape~\cite{yeom}, brings optimism on the possibility of achieving control over more and more particle features, including chirality, at the microscopic level. Computer simulations of hard particles will be helpful in the shape design of future building blocks.

\medskip

\section*{Methods}
The TTP studied here have $\gamma \leq \pi/3$ yielding $m=\pi\cos\gamma/(1+\cos\gamma)$ and $s=\pi\sin\gamma/(2+2\cos\gamma)$, such that the shape parameter $\nu=h(1+\cos\gamma)/\pi\cos\gamma - 2/\tan\gamma$ (when $\gamma \leq \pi/3$), where $h$ is the height of the particle. The volume of the particle is calculated by using standard formulas for orientable polyhedra that requires the knowledge of face normals and vertices positions. Overlaps between particles are detected by checking intersections between triangular faces using the RAPID library~\cite{rapid}.
Thousands of particles are simulated using standard MC simulation methods in either the $NVT$ or $NPT$ ensemble. In the former, MC moves consist in either single-particle translation or rotation whereas in the latter also volume-change moves (both isotropic and anisotropic scaling) are employed. Several millions of MC steps are performed both in the equilibration and production runs, where one MC step is defined as $N$ moves, with $N$ the number of particles. Different boundary conditions and initial configurations are used, as specified in the text. In the case of hard walls (located at $z=0$ and $z=L_z$), the overlap detection between particles and walls is performed by checking if any vertices of the polyhedra have coordinates $z<0$ or $z>L_z$. For the implementation of twisted boundary conditions we refer to~\cite{allen}. To obtain the equilibrium equation of state we combined results obtained by starting from an isotropic configuration, from a dilute lattice and from a dense aligned lattice (constructed by first obtaining the closest packing of a few particles in an orthogonal cuboidal box). States equilibrated at close pressures are also used as initial configurations to avoid kinetic traps. To determine the transitions between different thermodynamic phases, we have calculated several order parameters in both the $NPT$ and $NVT$ ensembles. Using equilibrated configurations we also set up long (more than $6*10^6$ MC steps) $NVT$ simulations to accurately measure the cholesteric pitch $\mathcal{P}$. After dividing the system in slabs, we compute the nematic director by diagonalizing the tensor  $\mathcal{Q}_{\mathbf{\hat{u}} \alpha \beta}=(\sum_{i} 3\mathbf{\hat{u}}_{i\alpha} \mathbf{\hat{u}}_{i\beta} - \delta_{\alpha\beta})/2n$, where $\alpha,\beta=x,y,z$ and $i=1,...,n$ with $n$ the number of particles in the slab. The same procedure is repeated for the medium particle axis $\mathbf{\hat{v}}$ and the short axis $\mathbf{\hat{w}}$. Note that we neglect here the polar nature of the particle as we assume  up-down symmetry $\mathbf{\hat{n}}_{\mathbf{\hat{w}}}=-\mathbf{\hat{n}}_{\mathbf{\hat{w}}}$. However, we checked that it did not affect our results. For each configuration we calculate $\cos(\theta(z)) \equiv \mathbf{\hat{n}}_{\mathbf{\hat{u}}}(z) \cdot \mathbf{\hat{n}}_{\mathbf{\hat{u}}}(z=0)$ and we bypass the up-down symmetry by taking the absolute value. After averaging hundreds of such profiles, we perform a one-parameter fit using $|\cos \left(2\pi z/\mathcal{P} \right)|$ to extract the cholesteric pitch $\mathcal{P}$. Although our procedure removes the (small) intrisinc drift of the system occuring over different configurations, the statistical error on $\mathcal{P}$ is still on the order of several $\omega$ as shown in Fig.~\ref{fig3}(c). As usual, the eigenvalues associated to the nematic directors are identified as nematic order parameters $S$. Averaging the profiles $S(z)$, the bulk values are obtained for each state points. The cholesteric helical structure and the sense of the twist are further confirmed by orientational pair-correlation functions as introduced in~\cite{memmer} and reported in the Supplementary Information.\\ 

The simulation results are compared with those obtained using a second-virial classical density functional theory that is extensively described in~\cite{belli,dussi}. The input of such theory is the pitch-dependent Legendre-expanded excluded-volume between two particles with orientation $\mathcal{R},\mathcal{R}'$ separated by a distance $\mathbf{r}$:
\begin{multline}
E_{ll'} (q) =  - \int d\mathbf{r} \, \oint d\mathcal{R} \, d\mathcal{R'} \\
\times f(\mathbf{r},\mathcal{R},\mathcal{R}') P_{l} (\mathbf{\hat{n}}_q (z) \cdot \mathbf{\hat{u}}) P_{l'} (\mathbf{\hat{n}}_q(0) \cdot \mathbf{\hat{u}}')
\end{multline}
with $P_l$ the normalized Legendre polynomial of grade $l$=$0,2,$...,$20$ (only even coefficients are considered), $q=2\pi/\mathcal{P}$ the chiral wave vector, $\mathbf{\hat{n}}_q (z) = \mathbf{\hat{x}} \, \sin qz +  \mathbf{\hat{y}} \, \cos qz $ the nematic director profile and $f$ the Mayer function that assumes a value $-1$ if particles overlap and $0$ otherwise. The coefficients $E_{ll'} (q)$ are calculated using a MC integration scheme. Once these coefficients are calculated, the orientation distribution function $\psi (\theta)$, with $\theta$ the polar angle with  respect to the local nematic director, is obtained by minimizing a Parsons-Lee-Onsager-like free-energy functional~\cite{onsager,parsons,lee} yielding the following equation:
\begin{multline}
\psi(\cos\theta)= \frac{1}{Z} \exp \left\{ -\rho \, G(\eta) \sum_{l,l'=0}^{\infty} \frac{E_{ll'} (q)}{4\pi^2} \right. \\
\left. \times \frac{1}{2} \left[ \mathcal{P}_l (\cos\theta) \psi_{l'} +\mathcal{P}_{l'} (\cos\theta) \psi_{l}  \right] \right\},
\end{multline}
with $\rho$ the number density, $G(\eta)$ the Parsons-Lee correction, $\psi_l$ the expansion coefficients of $\psi(\cos\theta)$ and $Z$ the normalization factor. Finally, the equilibrium pitch $\mathcal{P}$ is obtained by inserting back $\psi(\theta)$ into the functional and identifying the minimum of the free energy.

\section*{Acknowledgments}
We acknowledge financial support from a NWO-ECHO grant. We thank SURFsara (www.surfsara.nl) for the support in using the Lisa Computer Cluster. We thank Ren\'e van Roij, John Edison, Simone Belli, Laura Filion and Nick Tasios for useful discussions and/or critical comments on the manuscript.

\end{document}